\documentclass{article}

\usepackage{arxiv}

\usepackage[utf8]{inputenc} % allow utf-8 input
\usepackage[T1]{fontenc}    % use 8-bit T1 fonts
\usepackage{hyperref}       % hyperlinks
\usepackage{url}            % simple URL typesetting
\usepackage{booktabs}       % professional-quality tables
\usepackage{amsfonts}       % blackboard math symbols
\usepackage{nicefrac}       % compact symbols for 1/2, etc.
\usepackage{microtype}      % microtypography
\usepackage{graphicx}
\usepackage[round,authoryear]{natbib}
\usepackage{doi}
\usepackage{float}

\title{Support for AI Development:\\Automated Daily Measurement with Open Data and Code}
\author{ \href{https://orcid.org/0000-0002-4140-0268}{\includegraphics[scale=0.06]{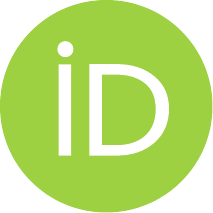}\hspace{1mm}Jason Jeffrey Jones}\thanks{\url{https://jasonjones.ninja} } \\
	Department of Sociology and Institute for Advanced Computational Science\\
	Stony Brook University\\
	Stony Brook, New York, USA \\
	\texttt{jason.j.jones@stonybrook.edu} \\
}

\renewcommand{\shorttitle}{Support for AI Development Measured Daily}

\hypersetup{
pdftitle={Support for AI Development - Automated Daily Measurement with Open Data and Code},
pdfauthor={Jason Jeffrey Jones},
}

\begin{document}
\maketitle

\begin{abstract}
This manuscript presents and advocates for a new form of scientific communication: free and open nowcasting of public opinion via web dashboard.  I present an open-source automated system that gathers new human responses to survey items daily, anonymizes and publicly distributes microdata, and presents analyses through a publicly viewable Web dashboard.  A demonstration implementation tracked support for further development of artificial intelligence among American adults.  As of 2026-05-31, the system had autonomously produced 766 daily estimates of support from ($N~=~8551$) respondents.  The findings underscore the need for continuous, high-frequency surveys to accurately track shifts in public opinion on transformative technologies like AI.  I argue that more scientists should adopt the method of open nowcasting, because it encourages transparency in research design and eases replication.
\end{abstract}

As social scientists, we should employ technology to make the process of discovering truth more efficient.  Here I present an open-source system with that aim.  Specifically, the system creates and automatically updates a new form of scientific communication: a daily nowcast of public opinion.

Nowcasting is defined as "predicting" the present, the very recent past and the very near future \citep{banbura_nowcasting_2011}.  As a practice, nowcasting began in meteorology, but more recently has been imported into social science.  Specifically, economists have nowcasted Gross Domestic Product (GDP) estimates \citep{bok_macroeconomic_2018} between official figures – which are low-frequency, delayed and often revised after initial release.  There is tremendous value in high-frequency, immediately available quantitative information, and that is just as true for American attitudes as it is for GDP figures.

Social science in general could benefit from more frequently updated data.  The General Social Survey (arguably the most widely used social science instrument) releases new data every two years.  The American National Election Studies project collects data every four years.  A successful precedent for {\itshape daily} data has been set, however.  The Global Database of Events, Language and Tone compiles scrapes of news stories into daily datasets released in real-time.  Researchers have found this useful; the work by \citet{leetaru_gdelt_2013} has been cited by over 1400 manuscripts.

Social science could also benefit from more open data.  Free and open datasets are widely available for academic studies of computer vision and natural language processing.  I believe it is not a coincidence these fields of study have made rapid, recent progress.  The availability of these datasets has been described as "central," "essential," and "crucial" by researchers \citep{emam_state_2021,lhoest_datasets_2021}.  I aim to bring open data to the study of public opinion for all the benefits it provides – including improved data quality, easing collaboration, standardized benchmarking, and increasing trust through transparent research practices.

The current work demonstrates an open-source system to collect survey response data at daily resolution and nowcast public opinion.  The goal is to provide research communities with a workflow that converts modest investments in time and money into consistently, persistently growing public datasets.

The focal item in the demonstration survey presented here was: How much do you agree with the statement: {\itshape I support further development of artificial intelligence}?  8551 total respondents indicated their agreement, explained their choice (optionally), and answered a few more questions.

The atypically high cadence --- 11 respondents surveyed every day --- provides uniquely useful temporal resolution.  The data provide insight on if, when, how and for whom sentiment toward AI has changed.

Several trends became clear.  These speak to the general public's changing relationship with AI; the results should inform and constrain theory and policy in the realm of AI and society.

\begin{itemize}
    \item Over time, support for further development of AI decreased.
    \item Political polarization on AI support emerged; support changed at different rates for Democrats and Republicans.
    \item Greater generalized trust and higher risk willingness were associated with more AI support.
    \item Male respondents reported higher AI support.  Older participants reported more support than younger --- more so the more recent the observation.
    \item Respondents expressed varied hopes, fears and uncertainties in their open-ended responses.  Compared to previous work, new categories of response appeared --- for example, concern regarding utilities consumption and appreciation of AI companionship.
\end{itemize}

The manuscript proceeds in the following manner.  Section 1 introduces and defines the Social Science Dashboard Inator system.  Section 2 motivates and discusses the demonstration project: estimating support for further development of artificial intelligence.  Section 3 details the methods of the demonstration project.  Section 4 presents the survey results, including analysis uniquely afforded by the daily data.  In Section 5, I review the present results, make predictions about future results, and discuss the limitations and advantages of the daily nowcasting dashboard approach.

\section{Social Science Dashboard Inator: A System to Collect, Distribute and Present Nowcasted Public Opinion}
The Social Science Dashboard Inator (SSDI, \url{https://github.com/jasonjeffreyjones/social-science-dashboard-inator}) is a free and open source system that automates this process:
\begin{enumerate}
    \item From a large pool of potential survey respondents, choose a small random set and distribute a short survey to them.
    \item Download survey responses, match these to known demographics of respondents and publish an anonymized, structured microdata file to a public data repository.
    \item Update a web dashboard that summarizes results and tests hypotheses.
\end{enumerate}

Using the system has many benefits.  The primary benefit is continuous, automated data collection.  A researcher invests the effort of a one-time setup and is then rewarded with new data every day.  Temporal trends are automatically detected in real-time.  

A secondary benefit is built-in commitment to research transparency.  The very best way to promote data sharing is to make it automatic and unavoidable.  Social Science Dashboard Inator \textit{forces} adherence to open-science practices.  Automated daily scripts collect data, share it publicly, and instantly conduct replicable analysis with open-source code.

A third benefit is flexibility.  The system may be adapted to address other social science research questions.  For example, another SSDI instantiation measures changes in self-concepts over time (Ipseity Daily, \url{https://jasonjones.ninja/social-science-dashboard-inator/ipseity-daily/}).  SSDI systems scale with funding investment.  If spending is increased, researchers can choose higher precision estimates and/or a larger battery of items.

Finally, SSDI systems create an affordance for short feedback loops: predict, observe, update models, repeat.  For these reasons, I believe automated daily nowcasting should be an aim more researchers pursue.

\subsection{Technical Specifications}
\label{subsec:tech-specs}
Table~\ref{tab:loop} illustrates the daily, looping, automated process.  Each day, new data is collected, wrangled, analyzed and publicly distributed.  Importantly, this process is completely automated.  A short series of open-source scripts (R and Python) perform each step in sequence.  The scripts are triggered by cron jobs and log their progress.  There is no manual process; daily data collection and analysis proceeds autonomously. The primary investigator and the broader community can watch as estimates become more precise and temporal trends appear in the compounding data.

\begin{table}
  \caption{Loop (daily, automated)}
  \label{tab:loop}
  \begin{tabular}{rl}
    \toprule
    Step&Event\\
    \midrule
    1 & New respondents recruited\\
    2 & New data collected\\
    3 & Cumulative file updated and shared\\
    4 & Analyses performed on new cumulative data\\
    5 & Web dashboard updated to display results\\
  \bottomrule
\end{tabular}
\end{table}

Table~\ref{tab:setup} illustrates the one-time manual setup process.  For social scientists, the setup contains familiar and unfamiliar steps.  Familiar to researchers will be creating surveys and paying for respondent recruitment.  Less familiar steps are forking a GitHub repository, creating \verb|cron| jobs and obtaining API credentials.  Adventurous researchers can learn these skills themselves.  Luckier researchers can use funding to pay technical staff.  In the near future, items can be piloted at no cost and with no setup through the Ryerson Project (\url{https://jasonjones.ninja/social-science-dashboard-inator/ryerson-project/about.html#faq})

An SSDI must host all materials and results in a public repository.  The repository contains the survey items and response options as text.  The repository should also contain all other files (e.g. R and Python scripts, HTML templates for the Web dashboard).  I encourage researchers to begin with the demonstration repository at \url{https://github.com/jasonjeffreyjones/social-science-dashboard-inator}.  The automated system continuously updates a publicly available cumulative data file.

\begin{table}
  \caption{Setup (one-time, manual)}
  \label{tab:setup}
  \begin{tabular}{rll}
    \toprule
    Step&Task&Recommended Tool\\
    \midrule
    1 & Compose an online survey& Qualtrics XM\\
    2 & Set up daily respondent recruitment& Prolific API, Python\\
    3 & Modify analysis code& R\\
    4 & Modify presentation templates& HTML, CSS\\
    5 & Set up web server& Institution or AWS\\
    6 & Upload all files& SFTP\\
    7 & Set up daily jobs& cron, command line\\
  \bottomrule
\end{tabular}
\end{table}

\section{Surveying Support for AI Development}

To pilot the Social Science Dashboard Inator, the author fielded a survey to American adults. The automated loop (Table~\ref{tab:loop}) delivered the survey to a random selection of 11 respondents from the approximately 90,000 eligible Prolific users every day.  The survey began running on April 18, 2024 and has run continuously since.  More than 8000 respondents have provided data.

\subsection{Why Survey Support for AI Development?}

Do Americans support further development of artificial intelligence (AI)?  It is common sense to quantify and persistently track a response to this question for several reasons.  First, many commercial entities are pursuing development of AI and making rapid progress that might cause disruptive change to economic and social systems.  Second, billions of US dollars are being spent by federal and state governments toward the aim of furthering development.  Third, there are outspoken sources which characterize further development of AI as a doomsday scenario that will end in human extinction \citep{yudkowsky_if_2025}.  Continuous observation of the distribution of Americans' support has practical application, and at the same time, presents an unprecedented opportunity to "nowcast" evolving public opinion on a developing technology.

There is reason to hypothesize that Americans are currently forming or revising their attitudes toward AI development.  Clear evidence demonstrates the salience of AI in media has recently increased \citep{ryazanov_how_2025}.  Search queries for artificial intelligence in 2025 were double the number they were five years previously \citep{google_trends_google_2025}.

Since the release and popular adoption of ChatGPT, it has appeared that AI development is in a moment of extreme acceleration.  I anticipated that incidents affecting public opinion both positively and negatively were predictably likely to happen but with unpredictable frequency and timing.  Continuous (rather than sporadic) measurement of AI support ensured that comparable data would be available before, during and after important developments.

\subsection{Previous Results}

Many careful studies of public opinion on AI have been conducted \citep{schepman_initial_2020,sindermann_assessing_2021,jones_attitudes_2024} .  Each presented data at annual or one-shot temporal resolution, however.  In the quickly-evolving domain of AI development, finer-grained data available at low-latency has the potential to be extremely valuable.

The present work was inspired by previous surveys.  Specifically, \citet{zhang_artificial_2019} posed the following question to a representative sample of American adults in 2018: {\itshape How much do you support or oppose the development of AI?}  They described their results this way: 

\begin{quote}
Americans express mixed support for the development of AI, although more support than oppose the development of AI ... A substantial minority (41\%) somewhat or strongly supports the development of AI.  A smaller minority (22\%) somewhat or strongly oppose its development.  Many express a neutral attitude: 28\% of respondents state that they neither support nor oppose while 10\% indicate they do not know.
\end{quote}

Following this lead, I surveyed Americans on the similar item \textit{How much do you oppose or support the development of Artificial Intelligence?} weekly from 2020 through 2022 \citep{jones_jones-skiena_2022}.  Response options were the values 1 to 7 with 1 labeled as \textit{Strongly oppose} and 7 \textit{Strongly support}.  This was done manually.  I spent every Monday morning preparing the next survey and updating the website with results from the previous week.  In my final update on 2022-04-29 I wrote: 

\begin{quote}
On average, the American public supports the development of Artificial Intelligence. In the data from April 14, 2022, the estimated mean response is 4.48, with a 95\% confidence interval of 4.12--4.83. On our 7-point scale ... a value of 4 indicates indifference, any value below that opposition and any point above support.  In 82 weeks of running this same survey item, we have never observed a mean estimate below the midpoint of the scale.
\end{quote}

Moving to a daily, autonomous system --- the present work --- was motivated by the idea that AI development was in a moment of extreme acceleration.  Continuous (rather than sporadic) measurement of AI support ensured that comparable data would be available before, during and after incidents as they developed.

\section{Survey Methods}

The focal item was \texttt{AI Support}, measured similarly to previous work \citep{jones_jones-skiena_2022,zhang_artificial_2019}.  Measures of \texttt{Generalized Trust} \citep{uslaner_moral_2002} and \texttt{Risk Willingness}  \citep{dohmen_individual_2011} were included.  The author speculated that less trusting individuals would imagine or foresee nefarious applications of AI and therefore display less AI Support.  The rapid development of AI is often portrayed as risky \citep{cave_scary_2019,yudkowsky_pausing_2023}, and therefore the author investigated the possibility of a relationship between Risk Willingness and AI Support.  Political party affiliation was queried by first asking respondents for their party affiliation and then asking toward which party they 'lean' for those who deny an affiliation \citep{pew_research_center_appendix_2014}.  Age and Sex were demographics made available for all respondents by the Prolific platform.

\subsection{Procedure}

The survey comprised six items, and the median completion time was just under two minutes.  Each respondent was paid \$0.30 and platform fees added another 33\%.  The survey was presented through Qualtrics XM online survey software.

Respondents were recruited through the Prolific Academic platform \citep{palan_prolific_2018}.  A randomly selected 11 new respondents were recruited each day.  (A previous reviewer asked why the survey cadence had not been hourly.  Another reviewer requested a power analysis justifying a daily sample of 11 respondents.  The author admires the idealism and/or research budgets of those reviewers.  The figure of 11 respondents per day was decided on to keep the cost of the survey reasonable for a self-funded lone researcher - about three dollars per day.)

Table~\ref{tab:demographics} presents the Age and Sex distribution of respondents.  The sample procedure provided good demographic coverage - although not a perfect representative sample - of the American adult population.

\begin{table}

\caption{Age and Sex Distribution}
\centering
\begin{tabular}[t]{lrrr}
\toprule
Age (Binned) & Female & Male & Unavailable\\
\midrule
18-24 & 506 & 439 & 12\\
25-34 & 1351 & 1218 & 20\\
35-44 & 1156 & 1074 & 6\\
45-54 & 848 & 623 & 8\\
55-64 & 516 & 317 & 1\\
%\addlinespace
65+ & 253 & 158 & 1\\
Unavailable & 1 & 0 & 43\\
\bottomrule
\end{tabular}
\label{tab:demographics}
\end{table}

This manuscript presents results from April 18, 2024 through May 31, 2026, but the automated scripts that recruit new respondents and process new data continue to run daily.  Updated analyses and the raw microdata are made publicly available through a Web dashboard: \url{https://jasonjones.ninja/social-science-dashboard-inator/jjjp-ai-daily-dashboard/}

%\begin{figure*}[!thb]
\begin{figure}[H]
  \centering
  \includegraphics[width=\textwidth]{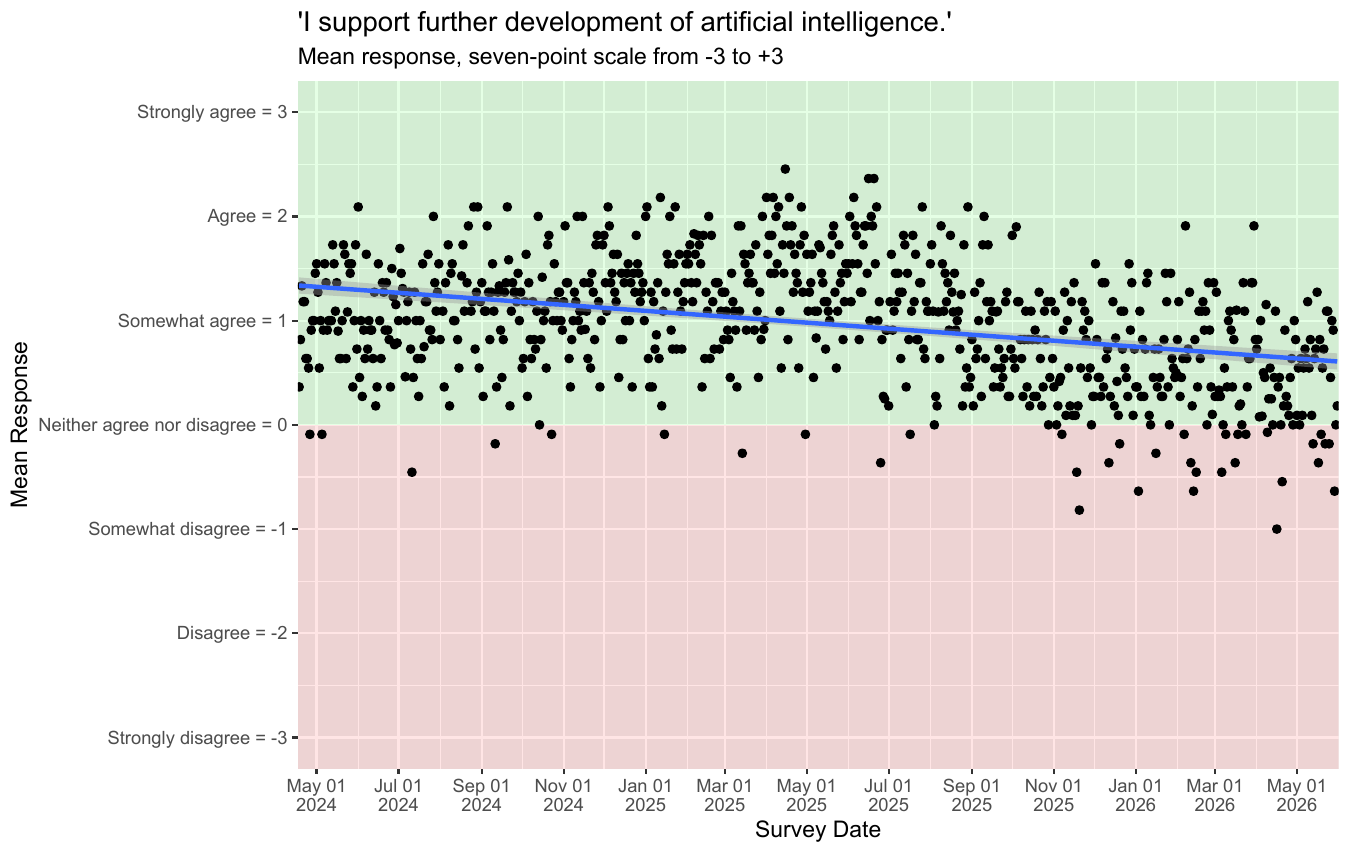}
  \caption{Each black point depicts a daily estimate of American adults' support for further development of AI. Specifically, each point was placed at the mean value of the day's agreement responses (mapped to numeric values).  \textbf{On average, support for further development of artificial intelligence decreased among American adults.} }
  \label{fig:daily-support-points}
%  \Description{On a plot, black points show American adults' support for further AI development on the vertical axis and dates on the horizontal axis.}
\end{figure}

\section{Results}

\subsection{Daily Support Series}

Daily estimates of American adults' support for further AI development are displayed in Figure~\ref{fig:daily-support-points}.  Daily data collection began in April of 2024, and support was modestly positive. After mapping the seven response options to the values -3 through +3, one finds a mean support value for the first 30 daily estimates of 0.99, with a 95\% confidence interval (CI) of [0.94,1.03].  For the last 30 daily estimates, the mean support value was 0.46, CI:[0.40,0.51].  In a linear regression over all daily estimates, the estimated coefficient for daily change was negative (-0.0009), and the confidence interval excluded zero [-0.0011,-0.0008].  Narratives of extreme change in AI support are contradicted by this evidence.

\subsection{Speculative Trend Break Analysis}

Due to budget constraint, each daily sample was small, and therefore estimates were noisy.  The lack of daily precision, however, has the positive tradeoff of high temporal resolution.  Of course, one can aggregate or smooth the individual estimates.  Doing so requires modelling decisions.  The author invites others to use the openly available microdata to model trends in their own way.  Below, I present the results based on my decisions.

\begin{figure}[h]
  \centering
  \includegraphics[width=\linewidth]{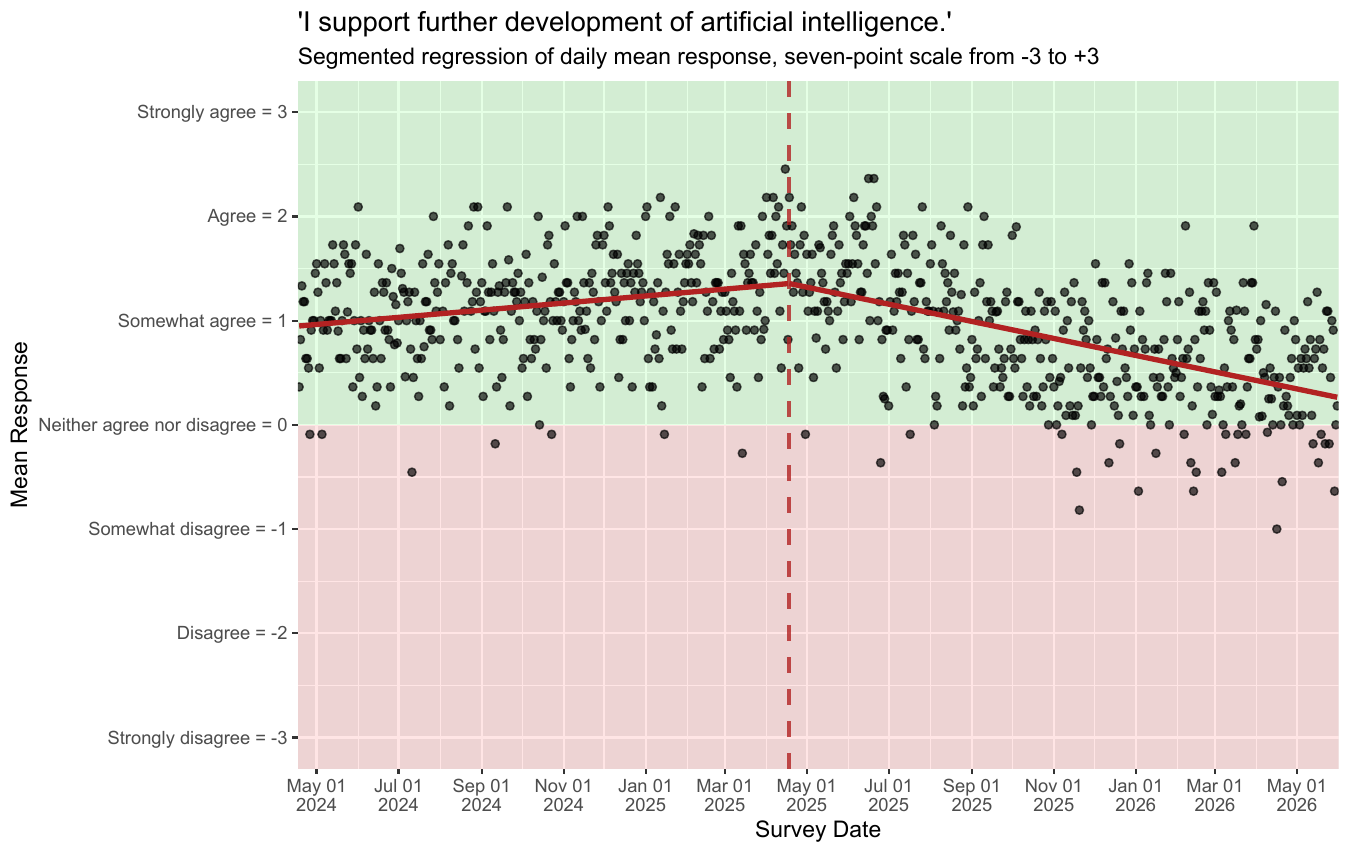}
  \caption{Segmented regression suggests a trend toward greater AI Support that reversed in 2025.}
  \label{fig:daily-support-segmented}
%  \Description{On a plot, two lines show American adults' support for further AI development increasing then decreasing.}
\end{figure}

Visually, the AI Support series appears to peak somewhere near mid-2025.  A Davies test \citep{davies_hypothesis_2002} confirms there is evidence of a slope change in the time series ($p<0.001$).  Segmented regression placed the breakpoint on April 18, 2025 (with a standard error of 19 days).  The slope leading up to that date had a statistically significant positive value, and the slope afterward is statistically significantly negative.  \textbf{Specifically, before the breakpoint, estimated support increased by approximately 0.41 scale points per year; afterward, it declined by approximately 0.98 scale points per year.}  The segmented model explained approximately 25.5\% of the variation in daily mean support, adjusted $R^2 = .255$.  Figure~\ref{fig:daily-support-segmented} visualizes this speculative two-epoch linear fit.

\subsection{Political Polarization}

\begin{figure*}[!htb]
  \centering
  \includegraphics[width=\linewidth]{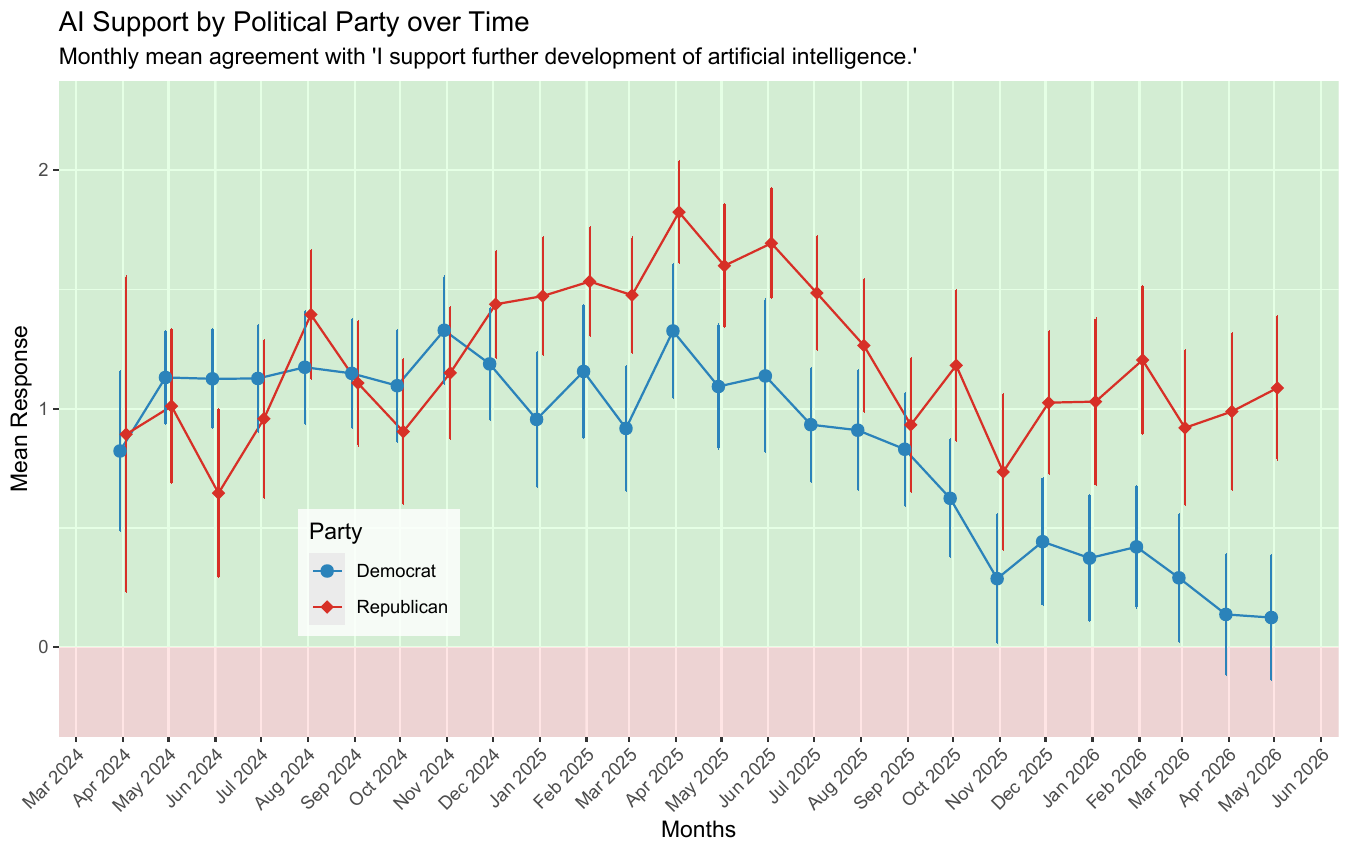}
  \caption{Support for further development of AI was plotted as a function of political party affiliation and time.  Responses were grouped by party and aggregated to months.  Points are mean values and bars represent 95\% confidence intervals.  \textbf{Republican support appears to exceed Democrat support beginning in December 2024.}  Note that the y-axis is truncated.  While there were many individual responses outside these bounds, no group means were in the truncated regions.}
  \label{fig:ai-polarization}
%  \Description{On a plot, lines and points show how American adults' support for further AI development differs between Democrats and Republicans over time.}
\end{figure*}

In the United States, surprisingly many preferences and attitudes are tied to political party affiliation (e.g. \cite{rogers_politicultural_2022}).  Topics of popular discussion become targets of political polarization (e.g. \cite{jones_evolution_2022}).  The author wished to explore these phenomena in relation to AI Support.  In Figure~\ref{fig:ai-polarization}, support is plotted as a function of {\itshape political party affiliation}.  To increase the precision of estimates, responses were aggregated to monthly resolution.

Beginning in December of 2024 and continuing after, observed mean Republican support for AI development exceeded Democrat support.  Noting that the Republican party won the Presidency and majorities in both houses of Congress in November of 2024, this pattern suggests an interesting hypothesis for further study. Partisans' support for powerful new technologies may depend upon the level of power for their preferred party in the federal government.

In a linear model, a negative coefficient for Month reached significance ($p<0.001$), as did the interaction of Month x Party ($p<0.001$). 

\subsection{Generalized Trust}

Individuals vary as to whether they are generally trusting or wary.  \citet{rosenberg_misanthropy_1956} operationalized Generalized Trust with the following item:

\begin{quote}Generally speaking, would you say that most people can be trusted or that you can’t be too careful in dealing with people?\end{quote}

This item is widely used to assess levels of generalized trust both within the United States in the General Social Survey \citep{davern_general_2023} and cross-nationally in the World Values Survey \citep{inglehart_world_2022}.

Trusting respondents reported higher AI Support (mean = 1.11) than Careful respondents (mean = 0.89).  A Welch two-sample $t$-test indicated a reliable difference between groups ($p < 0.001$), with the 95\% confidence interval for the mean difference ranging from 0.14 to 0.28. Results were confirmed using a Wilcoxon rank-sum test ($p < 0.001$), a nonparametric alternative
to the $t$-test that compares distributions rather than means.

\subsection{Risk Willingness}
To gauge Risk Willingness, respondents were asked:

\begin{quote}How do you see yourself: are you generally a person who is fully prepared to take risks or do you try to avoid taking risks? Please choose a number, where the value 0 means: ‘not at all willing to take risks’ and the value 10 means: ‘very willing to take risks’.\end{quote}

\begin{figure*}[!h]
  \centering
  \includegraphics[width=\linewidth]{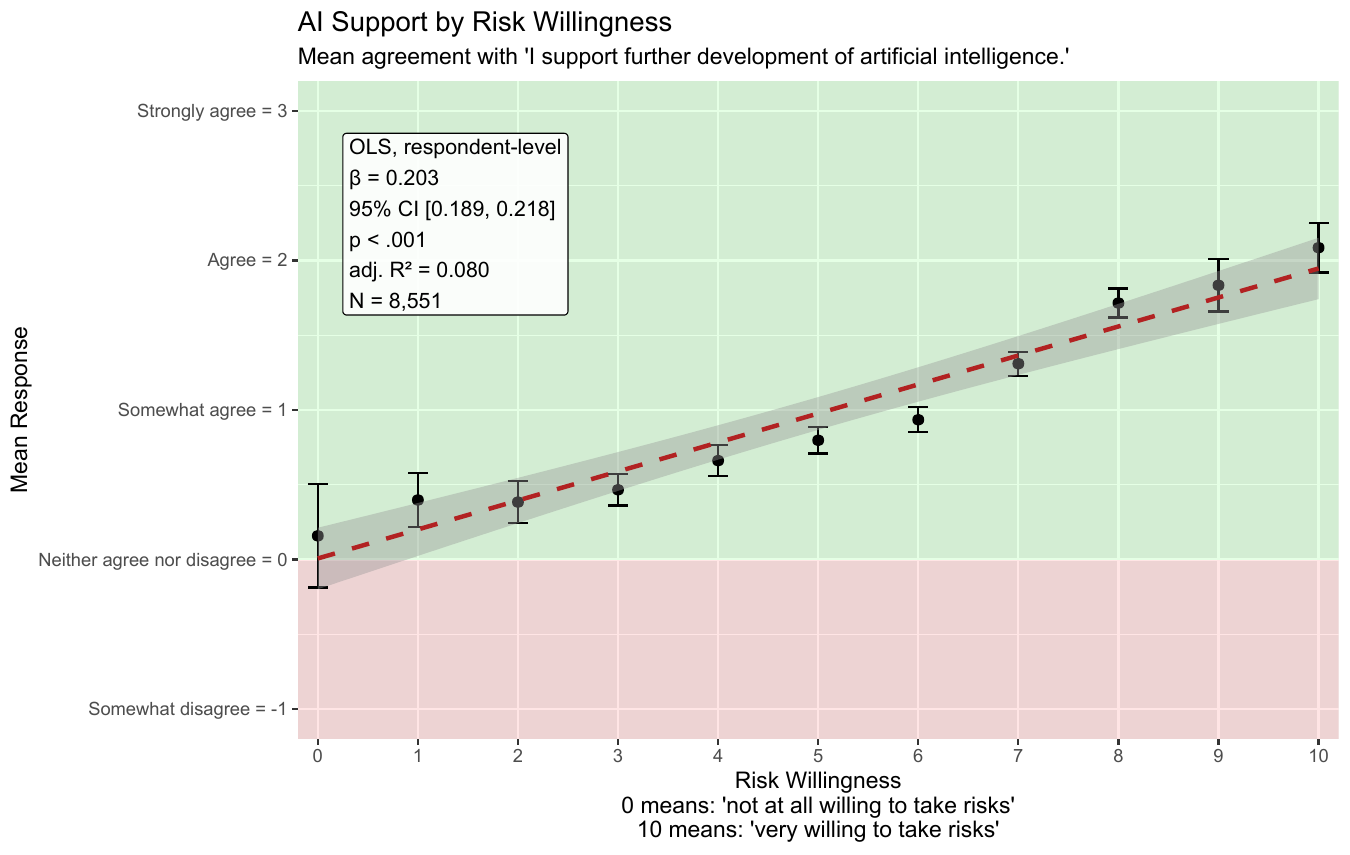}
  \caption{Support for further development of AI was plotted as a function of risk willingness.  Points are mean values and bars represent 95\% confidence intervals.  \textbf{Those less willing to take risks display weaker AI support.}  Note that the y-axis is truncated.  While there were many individual responses below \textit{Somewhat disagree}, no group means were in that region.}
  \label{fig:ai-risk}
%  \Description{On a plot, lines and points show how American adults' support for further AI development differs by their willingness to take risks.}
\end{figure*}

Called the {\itshape general risk question}, it has been claimed to "generate the best all-round predictor of risky behavior" \citep{dohmen_individual_2011}.  In Figure~\ref{fig:ai-risk}, AI support is plotted as a function of response to the {\itshape general risk question}.  A clear relationship emerged.  The greater a respondents' willingness to take risks, the greater their support for further development of AI ($p < 0.001$). Specifically, each one-point increase in risk willingness was associated with a $\hat{\beta} = 0.203$ increase in AI Support, 95\% CI [0.189, 0.218].

\subsubsection{Dynamics of the Support-Risk Relationship}

It is natural to wonder, will this relationship between risk willingness and AI support endure?  It might fade as AI becomes perceived as less risky, or the relationship might deepen (i.e. an even more positive slope in a future version of Figure~\ref{fig:ai-risk}).  With continual daily collection of data, an answer will inevitably become available. For the moment, we can examine what best explains the data so far.

The association between risk willingness and AI Support became slightly more positive over time, but the evidence was marginal. In a respondent-level model, the Risk Willingness $\times$ Time interaction was positive, $\hat{\beta} = 0.023$ per year, 95\% CI [-0.002, 0.048], $p = .067$.

Thus, the Risk-Support association was reliably positive throughout the study period, and there was only weak evidence that it sharpened over time.

\subsection{Age and Sex}

Several demographic variables were available from the Prolific respondent recruitment platform.  There was evidence for a statistically significant but small relationship between Age and Support.  In a respondent-level linear model, each additional decade of age was associated with a $\hat{\beta} = 0.04$ change in AI Support, 95\% CI [0.01, 0.07], $p = 0.007$.

\subsubsection{Dynamics of the Support-Age Relationship}

The temporal resolution of my data afford another analysis.  Simple observation of online discourse leads me to hypothesize that younger Americans are souring on AI more than older Americans.  We can estimate Support with an Age $\times$ Time interaction to test the hypothesis.

In a respondent-level linear model, Age was centered at 45 and measured in decades, and time was measured in years since daily sampling began. The Age $\times$ Time interaction was positive and statistically reliable, $\hat{\beta} = 0.092$ per decade per year, 95\% CI [0.047, 0.137], $p < .001$. This indicates that the \textbf{decline in support over time was steeper among younger respondents than among older respondents}. Equivalently, for each additional decade of age, the yearly decline in AI Support was estimated to be 0.092 scale points less negative. Model fit was modest, adjusted $R^2 = .018$.

%\begin{figure*}[!htb]
\begin{figure}[H]
  \centering
  \includegraphics[width=\linewidth]{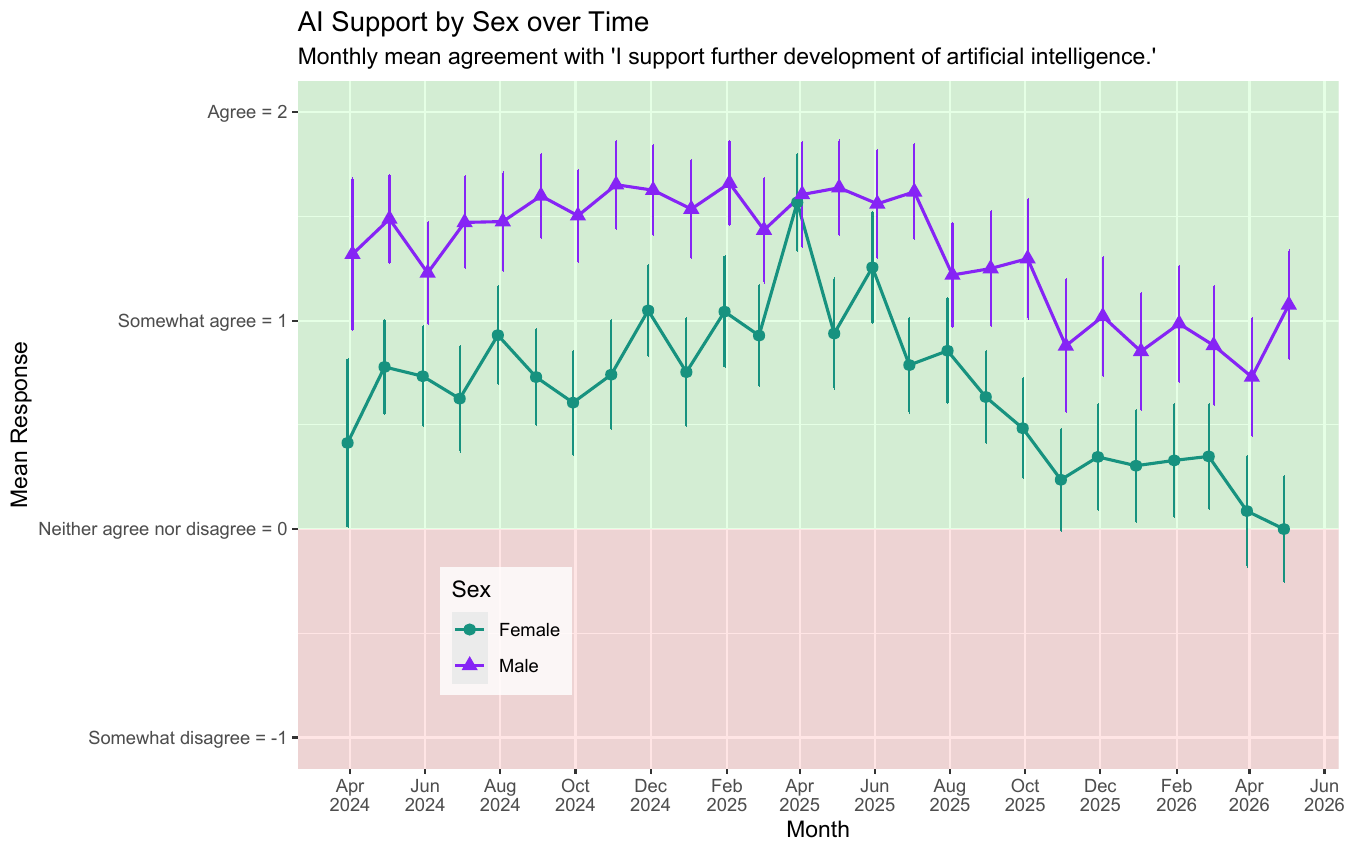}
  \caption{Support for further development of AI was plotted as a function of sex and time.  Those responding Female or Male were included and aggregated by month.  Points are mean values and bars represent 95\% confidence intervals.  \textbf{Males indicated greater AI Support.}  Note that the y-axis is truncated.  While there were many individual responses below \textit{Somewhat disagree}, no group means were in that region.}
  \label{fig:ai-sex}
\end{figure}

\subsubsection{Sex-Support Relationship}

Switching focus to Sex, Male respondents consistently reported higher AI Support than Female respondents, as Figure~\ref{fig:ai-sex} illustrates.  The overall mean for Male (1.34) was higher than Female (0.67)  A Welch two-sample $t$-test indicated a reliable difference between groups ($p < 0.001$), with the 95\% confidence interval for the mean difference ranging from 0.60 to 0.74. Results were confirmed using a Wilcoxon rank-sum test ($p < 0.001$).

\subsection{Open-Ended Text Responses}

In addition to daily numerical estimates of AI Support, the survey recorded a running stream of respondents' predictions, fears, and thoughts regarding AI.  After the primary AI Support item, respondents were invited to "Write 1 to 3 sentences explaining why."  5965 out of 8551 respondents provided a written rationale for their support or opposition.  All text considered here plus newly accumulating responses are available for download: \url{https://jasonjones.ninja/social-science-dashboard-inator/jjjp-ai-daily-dashboard/data/jjjp-ai-support-daily.csv}

In a survey of thousands of respondents across eight countries, \citet{kelley_exciting_2021} used open-ended responses to identify four emergent themes in public opinion of AI - exciting, useful, worrying, and futuristic. 
 These themes appeared again, as demonstrated below.

\subsubsection{Exciting}

\begin{quote}I think artificial intelligence is super interesting and cool. I would support further development of it.

\hfill---25 year old Female \textit{Strongly agree} on 2024-04-21\end{quote}

\begin{quote}It is helpful in so many ways and its potential is almost limitless.  It excites me to be honest.

\hfill---51 year old Female \textit{Strongly agree} on 2025-02-14\end{quote}

\subsubsection{Useful}

\begin{quote}Artificial intelligence is going to make the future easy and solve a lot of problems.

\hfill---22 year old Female \textit{Strongly agree} on 2025-04-18\end{quote}

\begin{quote}I see AI as a powerful tool for solving problems and improving productivity.

\hfill---33 year old Male \textit{Agree} on 2025-02-09\end{quote}

\subsubsection{Worrying}

\begin{quote}Do you want Skynet? Because that's how you get Skynet. We are playing with ideas and technology we are too dumb to understand, and implementing it before it's safe to use. There aren't even laws banning the development of a Skynet-like entity.

\hfill---32 year old Male \textit{Strongly Disagree} on 2024-04-21\end{quote}

\begin{quote}I think that the inherent development of artificial intelligence is not a bad thing; however, I worry that the capitalistic and exploitative nature of our society will use AI to further enslave the majority of the population into working for less or losing their jobs.

---25 year old Female \textit{Somewhat agree} on 2024-09-28\end{quote}

\subsubsection{Futuristic}

\begin{quote}Artificial intelligence could be a tectonic shift in how humans interact with the world. Further development could change everything about the human experience on Earth, the rewards greatly outweigh the risks.

\hfill---32 year old Male \textit{Agree} on 2024-04-30\end{quote}

\begin{quote}It's the future and we must embraced it

\hfill---53 year old Male \textit{Agree} on 2024-12-19\end{quote}

\subsubsection{New Directions}

Five years on from the work of \citet{kelley_exciting_2021}, it is appropriate some responses were less speculative and more concrete.  Dozens of respondents mentioned the environmental costs of AI and even more called for government regulation of the technology.  Some related individual experiences. For instance, two respondents reported AI had had negative personal economic impact:

\begin{quote}AI killed my career as a writer for one thing. Also, AI is a scammer's dream come true. Nothing that modern AI has done has been able to outweigh the potential for abuse that we've seen.

\hfill---42 year old Male \textit{Disagree} on 2024-05-03\end{quote}

\begin{quote}It is very obvious to me that AI will help humanity with certain tasks, but of course there are downsides that are already happening. (as a freelance artist, I have lost jobs to clients just creating stuff in AI instead of hiring me to create stuff by hand). Overall, I think it could lead to some pretty amazing things, but I think we should proceed with caution because we aren't exactly sure how this will end up 10-30 years from now.

\hfill---52 year old Male \textit{Somewhat agree} on 2024-09-15\end{quote}

Other respondents reported receiving or desiring a personal connection with AI:

\begin{quote}It helps my with my work. It makes my research work easier. It is like a personal friend

\hfill---19 year old Female \textit{Agree} on 2024-12-23\end{quote}

\begin{quote}I'm lonely and want an AI girlfriend

\hfill---35 year old Male \textit{Agree} on 2025-04-10\end{quote}

The four emergent themes of \citet{kelley_exciting_2021} will likely become inadequate as AI develops and becomes more prevalent within individuals' work and personal experiences.

%\begin{figure*}[!htb]
\begin{figure}[H]
  \centering
  \includegraphics[width=\linewidth]{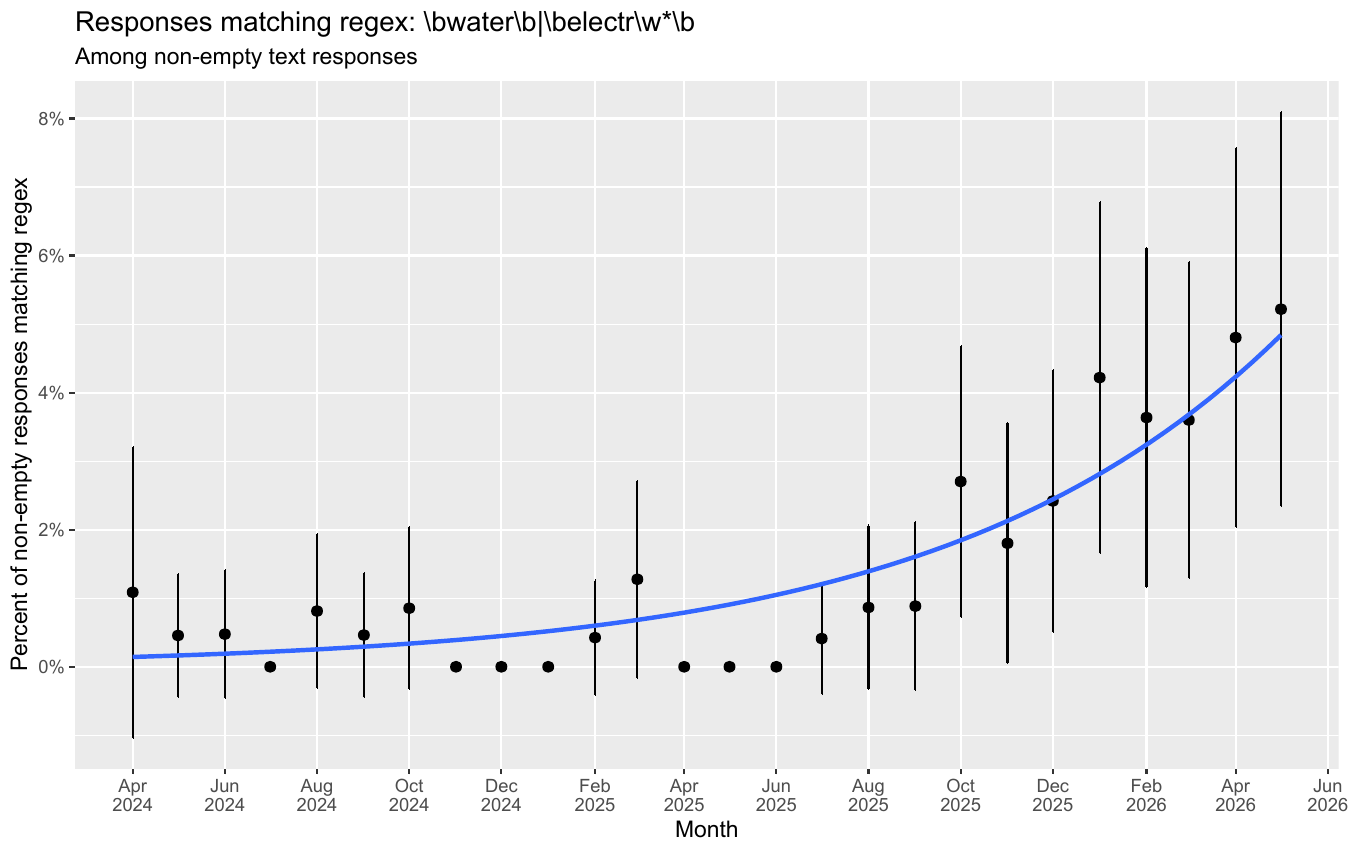}
  \caption{When asked to explain their level of AI Support, respondents increasingly mentioned water or electricity.}
  \label{fig:ai-rationale-water}
\end{figure}

To further illustrate, I tested a hunch that over time, there would be more mentions specifically of water and electricity in respondents' AI Support rationales.  As Figure~\ref{fig:ai-rationale-water} shows, an increasing percentage of text responses included the regular expression \texttt{\textbackslash bwater\textbackslash b|\textbackslash belectr\textbackslash w*\textbackslash b}.  Formally, in a logistic regression predicting whether a non-empty response matched the regular expression, each additional year was associated with substantially higher odds of a match, odds ratio = 6.49, 95\% CI [4.07, 10.80], $p < .001$.

\section{Discussion}

This manuscript has dual aims.  First, I have documented a system to estimate public opinion on a daily cadence.  Second, I have explored \textit{one} application: daily estimates of American's support for further development of artificial intelligence.  I will discuss the demonstration application first, and then conclude by advocating for more and broader Social Science Dashboard Inators.

\subsection{AI Support}

We learn a great deal from a simple survey, repeated often.  The rapid development of AI in 2024 onward has \textit{not yet} engendered a desire to pause AI development.  On average, American adults have moved little in their support for further AI development.  However, to the extent attitudes have changed, it has been toward decreasing support.

\subsubsection{AI Support associated with Trust and Risk}

This work is the first to explore the relationship between AI Support and the widely-studied constructs of Generalized Trust and Risk Willingness.  Remarkably, even those who considered themselves generally distrustful supported AI development, albeit slightly less than their trusting counterparts.  Future work should seek to explain how this could be so.  One answer could be that members of the public employed a dual process \citep{slovic_risk_2013} to gauge their AI Support.  In other words, they overcame initial feelings of discomfort around a new technology that many claimed was dangerous, and later came to believe through deliberation that the potential benefits would outweigh the potential risks.  Future studies should seek to test this explanation and others.

Researchers should consider adopting the 11-point Generalized Trust scale recommended by \citet{lundmark_measuring_2016}.  The dichotomous version used here (and recommended by \citet{uslaner_measuring_2015}) might have obscured a tighter relationship by compressing the available responses.  Gauging the relationship between \textit{institutional trust} \citep{marien_measuring_2011} (rather than interpersonal trust) and AI Support is another potentially more fruitful approach.  Some might subsume their distrust of individuals when considering how much to support a rapidly developing new technology and instead prioritize their trust in institutions such as courts and legislatures.  Much could be learned by direct comparison of different trust measures as predictors of AI Support.

Risk Willingness was reliably associated with AI Support ($\hat{\beta} = 0.203$, $p < 0.001$).  Consider Figure~\ref{fig:ai-risk}. Those very willing to take risks agree that AI development should move forward, while those not at all willing are equivocal.  This evidence comports with the ideas that AI is perceived as risky, and therefore more strongly supported by individuals more willing to take risks.  This result is strongly consistent with Diffusion of Innovation Theory \citep{rogers_diffusion_2003}, which explicitly positions risk tolerance as a key predictor of early adoption and support for emerging technologies. AI, currently perceived as innovative but with uncertain utility, may appeal disproportionately to individuals who generally embrace rather than avoid uncertainty.

It is beyond the scope of the current work, but researchers interested generally in trust and risk should make use of the compounding daily time series for these items.  This study of AI Support provides daily estimates of American adults' levels of trust and risk willingness as a side effect.  Event studies \citep{mackinlay_event_1997} could investigate whether these levels remain stable in the population or whether particular incidents shift estimates.

\subsubsection{Cleavages in AI Support}

Cleavages in AI Support emerged for sex and political party affiliation.  Both male and female respondents supported further development of AI, but male respondents slightly more. Both sexes participated in the downward trend late in the series.

The interaction evident in Figure~\ref{fig:ai-polarization} portends political polarization of AI support.  With the current data, one can only speculate why Republican AI support increased and overtook Democrat support.  Perhaps increasing commercialization of the technology was more appealing to Republicans.  Perhaps the increasing association of Elon Musk with AI was worrying to Democrats.  Perhaps if one's preferred party is in power, one sees more to be gained from developing technologies, and as the November election approached and passed, it became clear power would pass from Democrats to Republicans.  Regardless of cause, this development should be monitored closely.  AI may face a future resembling the politicization of climate science.

%Standardize previous results and combine.  Discuss.
%Is the increase small?
%Are the other effects small?

\subsubsection{Speculative Predictions for 2026}

The author believes - as many have argued \citep{hofman_prediction_2017,yarkoni_choosing_2017} - that a shift in focus to prediction would benefit the practice of social science.  One goal of the current work is to provide AI Support as a target variable for those willing to make such predictions.

Here I make predictions for 2026 based on intuition and speculation.  I invite others to make more precise and accurate predictions based on theory, statistical forecasting or their own judgment.

\textbf{AI Support will continue to decrease throughout 2026.}  In round numbers, as operationalized here, AI Support decreased from +1.5 to +0.5 over 25 months.  I assume the currently observed trend will simply continue, and thus, I predict the average AI Support in the survey will fall to near zero by the end of this year.  To make a precise prediction: mean AI Support will be +0.01 in December 2026 and turn negative in 2027.   Extrapolation such as this is perhaps ill-advised in the short term, and definitely nonsensical in the long-term.  However, this seems a sensible prediction for now given observed data.

\textbf{The relationships between Generalized Trust and Risk Willingness with AI Support will strengthen.}  I predict the $R^2$ of models predicting AI Support using only Generalized Trust or only Risk Willingness will be measurably higher in the next twelve months of data as compared to the previous twelve months data.  This is based simply on the intuition that AI is increasingly perceived as risky on its own and dangerous in untrusted hands.

\textbf{The relationships between demographics and AI Support will strengthen.}  It is a recurrent phenomenon that targets of popular discussion in the United States come to be age-, sex- and party-coded \citep{dellaposta_why_2015,mason_uncivil_2018}.  AI will be no different.  I predict the $R^2$ of models predicting AI Support using only Age, Sex, and Party Affiliation will be measurably higher in the next twelve months of data as compared to the previous twelve months.

Whether these predictions are proven accurate is less important than the fact that data exist such that they may be tested at all.  For this reason, and the following, data collection will continue through June 2027, at least.  More broadly, I agree with many others (e.g. \citet{floridi_ai4peopleethical_2018}): to develop AI in a manner to maximize human flourishing will require consistent, persistent measurement of public opinion.

\subsection{Dynamics of AI Support}

This study captured a turning point in AI Support.  On April 18, 2025 (standard error 19 days), Americans' increasing support for AI development ended, and a new regime of decreasing support began.  It is likely no other study has the temporal resolution to place that date.

More generally, measuring consistently and persistently affords analysis of temporal trends.  The opportunity extends beyond the first derivative (e.g. above) to second derivative and predictor-by-time evolution.  Per second derivative: time will tell whether the current decrease in support accelerates or slows.  Per predictor-by-time evolution: consider that Age already demonstrates an interaction with Time; a positive relationship between Age and Support only emerged late in the series.  The Risk-Support relationship is trending in a similar way; an existing positive relationship appears to be strengthening over time.  It is my opinion analysis that acknowledges relationship change over time is more honest and useful than the norm of ignoring time and the implied claim of eternal truth.

\subsection{Social Science Dashboard Inators Beyond AI Support}

It is the hope of the author that ongoing SSDIs create communities and communities create SSDIs.  The processes described in Section~\ref{subsec:tech-specs} are agnostic in regards to what attitudes are being nowcast.  Whatever a research community wishes to continuously monitor---be it fear, hope, sleep, affinity, polarization, etc.---an SSDI creates consistent data at high temporal resolution.  Time series data produced by nowcasting make it much easier to address questions of the form: When did X begin changing and why?

SSDIs are meant to tackle two crises facing social science.  First is the replication/reproducibility crisis.  Now in its second decade, the crisis began with a respected psychologist publishing a claim for extra-sensory perceptions of future events in a peer-reviewed journal using standard procedural and statistical techniques.  Despite much discussion about this absurd outcome, surprisingly little has changed in the way we conduct social science.

The second crisis is more recent and less discussed, but just as challenging.  Call this the "predictability crisis."  Growth in theories that successfully predict outcomes in human lives has not tracked in sync with the explosive growth in available data on human lives \citep{hofman_prediction_2017}.  As an example, consider the \cite{salganik_measuring_2020} mass collaboration.  The final report describes the unimpressive results of more than 100 researchers collaborating and competing on the common task of predicting life outcomes (such as grade point average) from extensive survey data starting at birth.  No matter the technique or predictive features teams focused on, prediction performance (measured with R-squared on holdout data) was rarely – and then only modestly – better than a baseline model.

Daily nowcasting with freely and openly available data and methods directly confronts both crises.  Open access to the data and analysis code makes reproducibility trivially easy.  Attempts at replication can begin with full knowledge of the procedures and materials used in the original SSDI.  In fact, there is nothing to prevent other researchers from \textbf{simultaneous} reproduction or replication.  Fundamental disagreements about the best way to measure a construct or conduct analysis on data can be addressed immediately by parallel dashboards.

SSDIs offer scientists an opportunity to undertake principled prediction.  Worthwhile theories and models should have something to say about the future trajectory of the daily time series.  Consistent, persistent estimates provide accountability to the predictions made by theories and models.  Collecting these estimates daily and continuously dramatically shortens the feedback loop as compared to a process of waiting years for data and further years for articles interpreting the data.

In summary, the author advocates for daily nowcasting as a powerful method to improve and accelerate social science.  Here I present an open source system to automate the process.  Further, I offer the initial results of the first demonstration project.  

Future development will streamline the setup process.  Ultimately, the goal is to provide research communities with tools that convert modest investments in time and money into consistently, persistently growing public datasets.

%Standardize previous results and combine.  Discuss.
%Is the increase small?
%Are the other effects small?

\subsection{Limitations}

The strength of the present work is consistent and persistent measurement, however, to achieve this meant sacrifices that led to limitations.  First, the sample was limited to the United States, and small respondent counts led to limited precision at the daily level.  A larger research budget would solve both problems.  The author only confined the sample to 11 Americans per day due to the variable cost of additional nations and subjects.

A future iteration would easily scale with more investment.  Greater precision in daily estimates would, of course, follow from larger daily samples.  Similarly, hourly estimates (requested by an anonymous reviewer) could be had for those so desiring.  For the moment, one must be satisfied with rough daily measurements, and rely on weekly, monthly or annual aggregation when more precision is necessary.

Second, the presently reported relationships may be unique to this moment or the American context.  For instance, one can imagine a future in which the benefits of automation disproportionately accrue to older adults, and young and old diverge much more sharply on AI Support.  As for the American context, the present data point to AI Support trajectories diverging by party affiliation, but cannot speak to whether and how this generalizes to politics outside the United States.

Third, the system relies on two paid services --- Qualtrics and Prolific.  These are familiar and available tools for social scientists \citep{douglas_data_2023,cushman_resource_2021}.  Both could conceivably be replaced with free and open tools, however.  Here they were included as convenient providers for survey delivery.  If necessary or desired, a different respondent recruitment tool (e.g. cloudresearch.com) could be substituted for Prolific.  Similarly, the Qualtrics dependency could be removed in favor of any other survey delivery module.  For example, in another SSDI instantiation (Ryerson Project, \url{https://jasonjones.ninja/social-science-dashboard-inator/ryerson-project/}), I removed Qualtrics completely by replacing it with custom open-source code.

\subsection{Conclusion}

Support for further development of AI is not static.  Consistent, persistent measurement at high temporal resolution is critically necessary if we wish to know if, when and how it is changing.  This work is a proof of concept.  Others may enrich our understanding by deploying a similar survey in other populations.  Even better, a funded, multinational collaborative could scale the present work into a fascinating continuous data stream.

Social Science Dashboard Inators are an alternative to traditional science publishing with many strong advantages.  Transparency is built in and unavoidable; all data generated instantly becomes a public good.  Source code is available so that flaws are quickly spotted and fixed.  Rather than static, hard-to-access files that spend years in "review" and then live behind paywalls, the publication is a publicly available website.  Most importantly, moving social science to dashboard publishing creates a norm of persistent and consistent measurement so that claims of truth are properly situated in temporal context.

\bibliographystyle{unsrtnat}
\bibliography{AI-Daily}

@inproceedings{leetaru_gdelt_2013,
	title = {Gdelt: {Global} data on events, location, and tone, 1979–2012},
	volume = {2},
	number = {4},
	booktitle = {{ISA} annual convention},
	publisher = {Citeseer},
	author = {Leetaru, Kalev and Schrodt, Philip A},
	year = {2013},
	pages = {1--49},
}

@incollection{banbura_nowcasting_2011,
	address = {Oxford},
	title = {Nowcasting},
	isbn = {978-0-19-539864-9},
	url = {https://www.oxfordhandbooks.com/view/10.1093/oxfordhb/9780195398649.001.0001/oxfordhb-9780195398649-e-8},
	language = {en},
	urldate = {2024-04-24},
	publisher = {Oxford University Press},
	author = {BańBura, M. and Giannone, D. and Reichlin, L.},
	editor = {Clements, M. P. and Hendry, D. F.},
	month = jul,
	year = {2011},
}

@article{jones_attitudes_2024,
	title = {Attitudes {Toward} {Artificial} {General} {Intelligence}: {Results} from {American} {Adults} in 2021 and 2023},
	copyright = {All rights reserved},
	issn = {27681254},
	shorttitle = {Attitudes {Toward} {Artificial} {General} {Intelligence}},
	url = {https://www.theseedsofscience.org/2024-attitudes-toward-artificial-general-intelligence},
	doi = {10.53975/8b8e-9e08},
	language = {en},
	urldate = {2024-02-20},
	journal = {Seeds of Science},
	author = {Jones, Jason Jeffrey and Skiena, Steven},
	month = feb,
	year = {2024},
}

@article{lundmark_measuring_2016,
	title = {Measuring generalized trust: {An} examination of question wording and the number of scale points},
	volume = {80},
	number = {1},
	journal = {Public Opinion Quarterly},
	publisher = {Oxford University Press US},
	author = {Lundmark, Sebastian and Gilljam, Mikael and Dahlberg, Stefan},
	year = {2016},
	pages = {26--43},
}

@article{schepman_initial_2020,
	title = {Initial validation of the general attitudes towards {Artificial} {Intelligence} {Scale}},
	volume = {1},
	journal = {Computers in human behavior reports},
	publisher = {Elsevier},
	author = {Schepman, Astrid and Rodway, Paul},
	year = {2020},
	pages = {100014},
}

@article{sindermann_assessing_2021,
	title = {Assessing the attitude towards artificial intelligence: {Introduction} of a short measure in {German}, {Chinese}, and {English} language},
	volume = {35},
	journal = {KI-Künstliche intelligenz},
	publisher = {Springer},
	author = {Sindermann, Cornelia and Sha, Peng and Zhou, Min and Wernicke, Jennifer and Schmitt, Helena S and Li, Mei and Sariyska, Rayna and Stavrou, Maria and Becker, Benjamin and Montag, Christian},
	year = {2021},
	pages = {109--118},
}

@article{zhang_artificial_2019,
	title = {Artificial intelligence: {American} attitudes and trends},
	journal = {Available at SSRN 3312874},
	author = {Zhang, Baobao and Dafoe, Allan},
	year = {2019},
}

@misc{yudkowsky_pausing_2023,
	title = {Pausing {AI} {Developments} {Isn}'t {Enough}. {We} {Need} to {Shut} it {All} {Down}},
	url = {https://time.com/6266923/ai-eliezer-yudkowsky-open-letter-not-enough/},
	language = {en},
	urldate = {2023-07-13},
	journal = {Time},
	author = {Yudkowsky, Eliezer},
	month = mar,
	year = {2023},
}

@article{bok_macroeconomic_2018,
	title = {Macroeconomic nowcasting and forecasting with big data},
	volume = {10},
	journal = {Annual Review of Economics},
	publisher = {Annual Reviews},
	author = {Bok, Brandyn and Caratelli, Daniele and Giannone, Domenico and Sbordone, Argia M and Tambalotti, Andrea},
	year = {2018},
	pages = {615--643},
}

@article{dellaposta_why_2015,
	title = {Why do liberals drink lattes?},
	volume = {120},
	number = {5},
	journal = {American Journal of Sociology},
	publisher = {University of Chicago Press Chicago, IL},
	author = {DellaPosta, Daniel and Shi, Yongren and Macy, Michael},
	year = {2015},
	pages = {1473--1511},
}

@book{mason_uncivil_2018,
	title = {Uncivil agreement: {How} politics became our identity},
	publisher = {University of Chicago Press},
	author = {Mason, Lilliana},
	year = {2018},
}

@article{yarkoni_choosing_2017,
	title = {Choosing prediction over explanation in psychology: {Lessons} from machine learning},
	volume = {12},
	number = {6},
	journal = {Perspectives on Psychological Science},
	author = {Yarkoni, Tal and Westfall, Jacob},
	year = {2017},
	note = {Publisher: Sage Publications Sage CA: Los Angeles, CA},
	pages = {1100--1122},
}

@book{rogers_diffusion_2003,
	title = {Diffusion of {Innovations}, 5th {Edition}},
	isbn = {978-0-7432-5823-4},
	language = {en},
	publisher = {Simon and Schuster},
	author = {Rogers, Everett M.},
	month = aug,
	year = {2003},
	note = {Google-Books-ID: 9U1K5LjUOwEC},
}

@incollection{slovic_risk_2013,
	title = {Risk as analysis and risk as feelings: {Some} thoughts about affect, reason, risk and rationality},
	booktitle = {The feeling of risk},
	publisher = {Routledge},
	author = {Slovic, Paul and Finucane, Melissa L and Peters, Ellen and MacGregor, Donald G},
	year = {2013},
	pages = {21--36},
}

@article{marien_measuring_2011,
	title = {Measuring political trust across time and space},
	journal = {Marien, S.(2011). Measuring Political Trust Across Time and Space. In: Hooghe M., Zmerli S.(Eds.), Political Trust. Why Context Matters},
	author = {Marien, Sofie},
	year = {2011},
	pages = {13--46},
}

@article{floridi_ai4peopleethical_2018,
	title = {{AI4People}—an ethical framework for a good {AI} society: opportunities, risks, principles, and recommendations},
	volume = {28},
	journal = {Minds and machines},
	publisher = {Springer},
	author = {Floridi, Luciano and Cowls, Josh and Beltrametti, Monica and Chatila, Raja and Chazerand, Patrice and Dignum, Virginia and Luetge, Christoph and Madelin, Robert and Pagallo, Ugo and Rossi, Francesca and {others}},
	year = {2018},
	pages = {689--707},
}

@article{mackinlay_event_1997,
	title = {Event studies in economics and finance},
	volume = {35},
	number = {1},
	journal = {Journal of economic literature},
	publisher = {JSTOR},
	author = {MacKinlay, A Craig},
	year = {1997},
	pages = {13--39},
}

@inproceedings{kelley_exciting_2021,
	title = {Exciting, useful, worrying, futuristic: {Public} perception of artificial intelligence in 8 countries},
	booktitle = {Proceedings of the 2021 {AAAI}/{ACM} {Conference} on {AI}, {Ethics}, and {Society}},
	author = {Kelley, Patrick Gage and Yang, Yongwei and Heldreth, Courtney and Moessner, Christopher and Sedley, Aaron and Kramm, Andreas and Newman, David T and Woodruff, Allison},
	year = {2021},
	pages = {627--637},
}

@article{uslaner_measuring_2015,
	title = {Measuring generalized trust: {In} defense of the ‘standard’question},
	journal = {Handbook of research methods on trust},
	publisher = {Edward Elgar Publishing},
	author = {Uslaner, Eric M},
	year = {2015},
	pages = {97--106},
}

@inproceedings{cave_scary_2019,
	title = {"{Scary} {Robots}" {Examining} {Public} {Responses} to {AI}},
	booktitle = {Proceedings of the 2019 {AAAI}/{ACM} {Conference} on {AI}, {Ethics}, and {Society}},
	author = {Cave, Stephen and Coughlan, Kate and Dihal, Kanta},
	year = {2019},
	pages = {331--337},
}

@misc{jones_evolution_2022,
	title = {The {Evolution} and {Polarization} of {Public} {Opinion} on {Vaccines}},
	url = {https://preprints.apsanet.org/engage/apsa/article-details/62013852a6fb4df4e24d9a3c},
	doi = {10.33774/apsa-2022-kxzt6-v2},
	language = {en},
	urldate = {2025-05-13},
	publisher = {APSA Preprints},
	author = {Jones, David and McDermott, Monika},
	month = feb,
	year = {2022},
}

@article{rogers_politicultural_2022,
	title = {Politicultural sorting: {Mapping} ideological differences in {American} leisure and consumption},
	volume = {50},
	number = {2},
	journal = {American Politics Research},
	author = {Rogers, Nick},
	year = {2022},
	note = {Publisher: SAGE Publications Sage CA: Los Angeles, CA},
	pages = {227--241},
}

@misc{pew_research_center_appendix_2014,
	title = {Appendix {B}: {Why} {We} {Include} {Leaners} {With} {Partisans}},
	shorttitle = {Appendix {B}},
	url = {https://www.pewresearch.org/politics/2014/06/12/appendix-b-why-we-include-leaners-with-partisans/},
	language = {en-US},
	urldate = {2025-05-09},
	journal = {Pew Research Center},
	author = {{Pew Research Center}},
	month = jun,
	year = {2014},
}

@article{ryazanov_how_2025,
	title = {How {ChatGPT} changed the media’s narratives on {AI}: a semi-automated narrative analysis through frame semantics},
	volume = {35},
	number = {1},
	journal = {Minds and Machines},
	publisher = {Springer},
	author = {Ryazanov, Igor and Öhman, Carl and Björklund, Johanna},
	year = {2025},
	pages = {1--24},
}

@misc{davern_general_2023,
	title = {General {Social} {Survey}, 2022},
	url = {https://osf.io/dmkaf/},
	doi = {10.17605/OSF.IO/DMKAF},
	publisher = {NORC at the University of Chicago},
	author = {Davern, Michael and Bautista, René and Freese, Jeremy and Herd, Pamela and Morgan, Stephen L.},
	month = oct,
	year = {2023},
}

@article{rosenberg_misanthropy_1956,
	title = {Misanthropy and {Political} {Ideology}},
	volume = {21},
	doi = {10.2307/2088418},
	number = {6},
	journal = {American Sociological Review},
	author = {Rosenberg, Morris},
	year = {1956},
	pages = {690--695},
}

@misc{inglehart_world_2022,
	title = {World {Values} {Survey}: {All} {Rounds} – {Country}-{Pooled} {Datafile} {Version} 3.0},
	url = {https://www.worldvaluessurvey.org/WVSDocumentationWVL.jsp},
	doi = {10.14281/18241.17},
	publisher = {JD Systems Institute \& WVSA Secretariat},
	author = {Inglehart, Ronald and Haerpfer, Christian and Moreno, Alejandro and Welzel, Christian and Kizilova, Kseniya and Diez-Medrano, Juan and Lagos, Marta and Norris, Pippa and Ponarin, Eduard and Puranen, Bi and Association, World Values Survey},
	editor = {Inglehart, Ronald and Haerpfer, Christian and Moreno, Alejandro and Welzel, Christian and Kizilova, Kseniya and Diez-Medrano, Juan and Lagos, Marta and Norris, Pippa and Ponarin, Eduard and Puranen, Bi},
	year = {2022},
	note = {Place: Madrid, Spain \& Vienna, Austria},
}

@misc{google_trends_google_2025,
	title = {Google {Trends}},
	url = {https://trends.google.com/trends/explore?date=today%205-y&q=artificial%20intelligence&hl=en},
	language = {en-US},
	urldate = {2025-04-30},
	journal = {Google Trends},
	author = {{Google Trends}},
	month = apr,
	year = {2025},
}

@misc{jones_jones-skiena_2022,
	title = {Jones-{Skiena} {Public} {Opinion} of {AI} {Dashboard}},
	url = {https://jasonjones.ninja/jones-skiena-public-opinion-of-ai/},
	urldate = {2024-11-20},
	author = {Jones, Jason Jeffrey and Skiena, Steven S},
	month = apr,
	year = {2022},
}

@article{dohmen_individual_2011,
	title = {Individual risk attitudes: {Measurement}, determinants, and behavioral consequences},
	volume = {9},
	number = {3},
	journal = {Journal of the european economic association},
	publisher = {Oxford University Press},
	author = {Dohmen, Thomas and Falk, Armin and Huffman, David and Sunde, Uwe and Schupp, Jürgen and Wagner, Gert G},
	year = {2011},
	pages = {522--550},
}

@article{palan_prolific_2018,
	title = {Prolific. ac—{A} subject pool for online experiments},
	volume = {17},
	journal = {Journal of behavioral and experimental finance},
	publisher = {Elsevier},
	author = {Palan, Stefan and Schitter, Christian},
	year = {2018},
	pages = {22--27},
}

@article{emam_state_2021,
	title = {On the state of data in computer vision: {Human} annotations remain indispensable for developing deep learning models},
	journal = {arXiv preprint arXiv:2108.00114},
	author = {Emam, Zeyad and Kondrich, Andrew and Harrison, Sasha and Lau, Felix and Wang, Yushi and Kim, Aerin and Branson, Elliot},
	year = {2021},
}

@article{lhoest_datasets_2021,
	title = {Datasets: {A} community library for natural language processing},
	journal = {arXiv preprint arXiv:2109.02846},
	author = {Lhoest, Quentin and Del Moral, Albert Villanova and Jernite, Yacine and Thakur, Abhishek and Von Platen, Patrick and Patil, Suraj and Chaumond, Julien and Drame, Mariama and Plu, Julien and Tunstall, Lewis and {others}},
	year = {2021},
}

@book{uslaner_moral_2002,
	address = {New York, NY, USA},
	title = {The moral foundations of trust},
	publisher = {Cambridge University Press},
	author = {Uslaner, Eric M},
	year = {2002},
}

@article{hofman_prediction_2017,
	title = {Prediction and explanation in social systems},
	volume = {355},
	url = {https://www.science.org/doi/abs/10.1126/science.aal3856},
	doi = {10.1126/science.aal3856},
	number = {6324},
	urldate = {2021-10-01},
	journal = {Science},
	publisher = {American Association for the Advancement of Science},
	author = {Hofman, Jake M. and Sharma, Amit and Watts, Duncan J.},
	month = feb,
	year = {2017},
	pages = {486--488},
}

@article{salganik_measuring_2020,
	chapter = {Social Sciences},
	title = {Measuring the predictability of life outcomes with a scientific mass collaboration},
	volume = {117},
	copyright = {Copyright © 2020 the Author(s). Published by PNAS.. https://creativecommons.org/licenses/by-nc-nd/4.0/This open access article is distributed under Creative Commons Attribution-NonCommercial-NoDerivatives License 4.0 (CC BY-NC-ND).},
	issn = {0027-8424, 1091-6490},
	url = {https://www.pnas.org/content/117/15/8398},
	doi = {10.1073/pnas.1915006117},
	language = {en},
	number = {15},
	urldate = {2021-10-01},
	journal = {Proceedings of the National Academy of Sciences},
	publisher = {National Academy of Sciences},
	author = {Salganik, Matthew J. and Lundberg, Ian and Kindel, Alexander T. and Ahearn, Caitlin E. and Al-Ghoneim, Khaled and Almaatouq, Abdullah and Altschul, Drew M. and Brand, Jennie E. and Carnegie, Nicole Bohme and Compton, Ryan James and Datta, Debanjan and Davidson, Thomas and Filippova, Anna and Gilroy, Connor and Goode, Brian J. and Jahani, Eaman and Kashyap, Ridhi and Kirchner, Antje and McKay, Stephen and Morgan, Allison C. and Pentland, Alex and Polimis, Kivan and Raes, Louis and Rigobon, Daniel E. and Roberts, Claudia V. and Stanescu, Diana M. and Suhara, Yoshihiko and Usmani, Adaner and Wang, Erik H. and Adem, Muna and Alhajri, Abdulla and AlShebli, Bedoor and Amin, Redwane and Amos, Ryan B. and Argyle, Lisa P. and Baer-Bositis, Livia and Büchi, Moritz and Chung, Bo-Ryehn and Eggert, William and Faletto, Gregory and Fan, Zhilin and Freese, Jeremy and Gadgil, Tejomay and Gagné, Josh and Gao, Yue and Halpern-Manners, Andrew and Hashim, Sonia P. and Hausen, Sonia and He, Guanhua and Higuera, Kimberly and Hogan, Bernie and Horwitz, Ilana M. and Hummel, Lisa M. and Jain, Naman and Jin, Kun and Jurgens, David and Kaminski, Patrick and Karapetyan, Areg and Kim, E. H. and Leizman, Ben and Liu, Naijia and Möser, Malte and Mack, Andrew E. and Mahajan, Mayank and Mandell, Noah and Marahrens, Helge and Mercado-Garcia, Diana and Mocz, Viola and Mueller-Gastell, Katariina and Musse, Ahmed and Niu, Qiankun and Nowak, William and Omidvar, Hamidreza and Or, Andrew and Ouyang, Karen and Pinto, Katy M. and Porter, Ethan and Porter, Kristin E. and Qian, Crystal and Rauf, Tamkinat and Sargsyan, Anahit and Schaffner, Thomas and Schnabel, Landon and Schonfeld, Bryan and Sender, Ben and Tang, Jonathan D. and Tsurkov, Emma and Loon, Austin van and Varol, Onur and Wang, Xiafei and Wang, Zhi and Wang, Julia and Wang, Flora and Weissman, Samantha and Whitaker, Kirstie and Wolters, Maria K. and Woon, Wei Lee and Wu, James and Wu, Catherine and Yang, Kengran and Yin, Jingwen and Zhao, Bingyu and Zhu, Chenyun and Brooks-Gunn, Jeanne and Engelhardt, Barbara E. and Hardt, Moritz and Knox, Dean and Levy, Karen and Narayanan, Arvind and Stewart, Brandon M. and Watts, Duncan J. and McLanahan, Sara},
	month = apr,
	year = {2020},
	pages = {8398--8403},
}

@book{yudkowsky_if_2025,
	address = {New York},
	title = {If {Anyone} {Builds} {It}, {Everyone} {Dies}: {Why} {Superhuman} {AI} {Would} {Kill} {Us} {All}},
	isbn = {978-0-316-59564-3},
	shorttitle = {If {Anyone} {Builds} {It}, {Everyone} {Dies}},
	language = {English},
	publisher = {Little, Brown and Company},
	author = {Yudkowsky, Eliezer and Soares, Nate},
	year = {2025},
}

@article{douglas_data_2023,
	title = {Data quality in online human-subjects research: {Comparisons} between {MTurk}, {Prolific}, {CloudResearch}, {Qualtrics}, and {SONA}},
	volume = {18},
	number = {3},
	journal = {Plos one},
	publisher = {Public Library of Science San Francisco, CA USA},
	author = {Douglas, Benjamin D and Ewell, Patrick J and Brauer, Markus},
	year = {2023},
	pages = {e0279720},
}

@article{davies_hypothesis_2002,
	title = {Hypothesis testing when a nuisance parameter is present only under the alternative: linear model case},
	journal = {Biometrika},
	publisher = {JSTOR},
	author = {Davies, Robert B},
	year = {2002},
	pages = {484--489},
}

@article{cushman_resource_2021,
	title = {Resource review—using qualtrics core {XM} for surveying youth},
	volume = {16},
	number = {1},
	journal = {Journal of Youth Development},
	author = {Cushman, Jennifer and Kelly, Miriah Russo and Fusco-Rollin, Maryann and Faulkner, Ryan},
	year = {2021},
	pages = {5},
}

\end{document}